\documentclass{appolb}
\usepackage{epsfig}

\begin{document}
\preprint{}
\title{Wounded Nucleons, Wounded Quarks, and Relativistic Ion 
Collisions%
\thanks{Presented at Cracow School of Theoretical Physics, XLVI 
Course, Zakopane, 2006}%
}
\author{Helena Bia{\l}kowska
\address{So{\l}tan Institute for Nuclear Studies, Warsaw, Poland}
}
\maketitle
\begin{abstract}

A concept of  wounded nucleons and/or wounded quarks plays an important 
role in parametrizing 
and to some extent explaining many a feature of the relativistic ion 
collisions. This will be illustrated in a historical perspective, up to 
and including the latest developpments. \end{abstract}
  
\section{Hadron - nucleus collisions}

Thirty years ago Andrzej Bia{\l}as has introduced a concept of a wounded 
nucleon, that is, a nucleon that has interacted at least once.
The Wounded Nucleon Model \cite{ABWound} - as usual - started from 
experimental observations, concerning high energy hadron-nucleus 
interactions (see Fig1). A series of Fermilab experiments \cite{Fermilab}, 
an 
European NA5 experiment \cite{NA5} and lots of emulsion data 
 have led to a somewhat surprising regularity: the average 
charged particle multiplicity increases in such collisions more slowly 
than the number of individual nucleon-nucleon collisions. This number, 
denoted usually by $\nu$, is given by
\begin{equation}
\nu=A \sigma_{hp}/\sigma_{hA},
\end{equation}
and the data gives the following dependence of the ratio of charged 
particles produced in hadron-nucleus collision to that for hadron-proton 
collision:
\begin{equation}
R=\left< n \right>_{hA} / \left< n \right>_{hp} = (1 + \nu) / 2.
\end{equation}
This is just the ratio of participants in hadron-nucleus (1 from hadron 
and $\nu$ from nucleus of mass A) and hadron-proton (2).
The Wounded Nucleon Model, WNM, states, that particle production in a 
nuclear collision is  a superposition of independent contributions 
from the wounded nucleons in the projectile and target. Thus one can just 
measure particle production in elementary collisions, count the 
wounded/participating nucleons in the target, and obtain the total 
particle multiplicity in hadron-nucleus collision. This is a rather strong 
statement, as one could expect that after each new hit, a nucleon would be 
less prone to produce new particles.

In Fig2  \cite{PhobosAGS} we observe a remarkable success of the model.
The ratio of total charged particle multiplicity from hadron-nucleus 
collisions, normalized to this multiplicity in elementary collisions, 
remains directly proportional to half the number of participants. This 
holds both for AGS energy range and RHIC data on d-Au at 200 GeV/c. Note 
that not only we have the proportionality to the number of participants, 
but also the scaling with elementary collisions data.

At the time when new experiments on nucleus-nucleus collisions were 
planned, new ideas appeared, originating from the wounded objects concept. 
Andrzej Bia{\l}as et al. (\cite {AB77}) and Vladimir Anisovitch et al. 
(\cite{Anis78}) have suggested, that it is rather the number of wounded 
quarks in the colliding objects that determines the produced particle 
multiplicity. Thus the ratio of particle multiplicity in nucleus-nucleus 
collision to that in elementary collision would be given by
\begin{equation}
R_{AB} = \frac{\nu_{AB}}{\nu_{qA} \nu_{qB}}.
\end{equation}
From the concept of wounded quarks to specific model predictions - it can 
be a long way. One can calculate the number of wounded quarks assuming a 
hypothetical value of quark-quark cross section, and a spatial 
distribution of nuclear matter - and hence quarks - in the colliding 
objects. The second step, particle production from a wounded quark, is 
open for many assumptions.

In 1982 Bia{\l}as \cite{Bialas1982} has given specific predictions for 
nuclear collisions, based on the Additive Quark Model, AQM. In this model, 
particle production would originate from three sources: breaking of the 
color strings between quarks from the projectile and target, fragmentation 
of the wounded quarks, and fragmentation of the spectator quarks.

I cannot refrain here from quoting one of the first nucleus-nucleus 
collision data, compared to `wounded nucleons' and `wounded quarks' model 
predictions. These were the results of a series of JINR Dubna experiments 
with proton, deuteron, alpha and carbon beams at 4.2 GeV/N (the highest 
energy nuclear beams at the time), incident on tantalum target 
\cite{Dubna}.The following table 
gives a comparison of particle multiplicities with the AQM.

With the first data on particle production from really high energy 
nucleus - nucleus collisions, an attempt at parametrizing produced particle 
multiplicities in terms of wounded objects was undertaken by K.Kadija et 
al., \cite{Kreso}. They have shown a consistent parametrization of 
production rates of negative hadrons - proportional to the number of 
wounded nucleons, and K0 - proportional to the number of wounded quarks - 
as shown in Fig.3. 
This was at the epoch when the highest energy was CERN SPS 200GeV/N, and 
the `heavy ion' was an oxygen, and at most - sulphur.

With the advent of higher statistics, higher energy and  higher mass 
numbers of colliding nuclei, such simple parametrizations are not common. 
Yet a `wounded 
object' concept has some very interesting comebacks.

Based on the data from RHIC,  $\sqrt{s}$ 200 GeV/N d - Au, a new version of 
the WNM was proposed by Bia{\l}as and Czy{\.z} \cite{ABdAu} (actually, 
this was presented for the first time at the Zakopane School 2 years 
ago!). The basic assumption is that particle production can be represented 
as a superposition of independent contributions from wounded nucleons in 
the projectile and the target. This applies not only to the total charged 
particle multiplicity, but to the longitudinal spectra as well. The 
density of particles in nucleus A - nucleus B collision is given by
\begin{equation}
\frac{dN_{AB}}{dy} = w_A F_A(y) + w_B F_B(y)
\end{equation}
and the model requires
\begin{equation}
F_B(y) = F_A(-y)
\end{equation}
(F is a contribution from a single wounded nucleon).

The first consequence is that
\begin{equation}
R_{AB}(y = 0) = \frac12 (w_A + w_B)
\end{equation}
This is very well checked by RHIC data, as shown in Fig.4.

For the full 
(pseudo) rapidity range, the authors construct symmetric and antisymmetric 
component:
\begin{equation}
G(\eta) = \frac{dN(\eta)}{d\eta} \pm \frac{dN(-\eta)}{d\eta}
\end{equation}
and compare to data on symmetric and antisymmetric part of spectra for 
several centralities of d-Au, as measured by the PHOBOS experiment at 
RHIC. This is illustrated in Fig.5. Given rather large experimental 
uncertainties, the parametrization of multiplicities and spectra is very 
reasonable. The authors stress that the contribution from one wounded 
nucleon extends over almost full rapidity range.

An interpretetion is given in a paper by Bia{\l}as and Je{\.z}abek 
\cite{AB+Jez}. The authors propose a two step particle production: 
multiple color exchanges between partons from projectile and target, and 
subsequent particle production from color sources created in the first 
step. In authors interpretation, a reasonable description of data by the 
model implies some sort of saturation -  the number of color sources per 
unit rapidity is independent  of the number of color exchanges between the 
projectile and target.

Thus for global characteristics of particle production in hadron - nucleus 
collisions, the WNM works surprisingly well. One should stress, that this 
applies to the total charged particle multiplicities. A more differential 
study of identified particles, such as strangeness carrying mesons and 
baryons, can not be described as a simple superposition of hadron - 
nucleon collisions \cite{Tania}.
 
\section{Nucleus - nucleus collisions}

For the `true' nuclear collisions, when two heavy nuclei collide, the 
Wounded Nucleon Model does not work. This is clearly seen from Fig.6 
\cite{PhobWhite}, showing total charged particle multiplicity per 
participant 
pair for several centralities of Au - Au collisions at four RHIC 
energies. To the first approximation,  thus normalized  multiplicity is 
flat as a 
function of the number of participants - but it clearly exceeds the 
multiplicity from proton - proton collisions.

Still, the proportionality of these multiplicities to the number of 
participants holds. Various approaches to specific choice of `effective 
energy' for proton- proton collisions were used. In Fig.7 we see a clear 
proportionality of nuclear multiplicities normalized to proton - proton 
multiplicity to the number of participants, both for Au - Au and d - Au 
collisions - but the protonic multiplicity is taken at twice the Au energy 
per nucleon \cite{PhobWhite}. This supposedly accounts for the 
leading baryon effect. Fig.8 again shows this proportionality - but 
with Au - Au data normalized to the multiplicity in e+ - e- collisions 
\cite{Busza}; an apparently unexpected universality in particle 
production from vastly different objects.

The surprising scaling of particle production in nuclear collisions with 
the number of participating nucleons extends to other characteristics, 
such as rapidity spectra and even transverse distributions. Fig.9 shows a 
comparison of pseudorapidity distributions for Cu - Cu and Au - Au 
collisions, measured for the same number of participating nucleons, at 
62.4 and 200 GeV/N, by the Phobos Collaboration \cite{Phobrap}, with the 
distributions practically identical. This geometric scaling works also for 
transverse spectra, and, surprisingly enough, extends even for transverse 
momenta as high as 6 GeV/c, as illustrated in Fig.10 \cite{Phobpt}.

There is a recent revival of the wounded quark parametrization ideas. 
Eremin and Voloshin \cite{Erem} draw the attention to the fact that 
midrapidity density of charged particle production in nuclear collisions, 
normalized to p - p data, show an increase with the number of nucleon 
participants, as seen in Fig.11. The authors calculate both numbers: of 
nucleon and quark 
participants, 
using Nuclear Overlap Model of K.Eskola et.al \cite{Eskola} - see the 
illustration in Fig.12. Then they parametrize the RHIC Au - Au results for 
midrapidity particle density  with the calculated participant numbers. As 
seen from Fig.13, scaling by quark participants flattens out the 
centrality dependence - this results from relative increase in the number 
of interacting constituent quarks in more central collisions.

Netrakanti and Mohanty \cite{Netrak} have applied the same idea also to 
the SPS 
data at $\sqrt{s}$ 17.2 GeV/c on charged particles and $\gamma$  
production in 
Pb - Pb collisions (WA98), illustrated in Fig.14, and Bhaskar De and 
S.Bhattacharyya \cite{De} to 
the NA49 data on identified particle production. They claim to observe a 
better and more unified description of midrapidity density yields for 
various secondaries with the quark participant picture. A word of caution 
is in order here. The treatment of 
the NA49 data is somehow inconsistent, the authors mixing the integrated 
yields with midrapidity yields for different particles. In a forthcoming 
study \cite{My} the data from NA49 will be consistently compared with 
constituent quark scaling.

 Let it be stated here, that a detailed study 
of strange particle production in light and heavy ion collisions 
\cite{Claudia} has definitely ruled out a simple participant scaling.

An energy dependence of particle production in nuclear collisions, 
compared to the production in elementary collisions, is a subject of 
another attempt at a description in terms of nucleon vs quark 
participants. The author, R.Nouicer \cite{Nouicer} looks at the energy 
dependence of particle density per  participant pair in central 
nucleus - nucleus collisions  vs the same quantity in proton - proton 
collisions. In terms of nucleon participants, the two sets of data follow 
distinct lines, as seen from Fig.15. With the normalization to the 
participant constituent quarks, both lines coincide. Yet a  very 
debatable point arises. As first pointed out by Barbara Wosiek, the author 
normalizes proton - proton data by the number of quark participants for 
`most central' collisions (while the pp data comprize all, minimum bias 
interactions). Normalization by `minimum bias' number of 
quark participants disturbs the common trend of data, as indicated by 
large points in Fig.15.

\section {Conclusions}
In summary, the very idea of particle production 
originating from 
elementary constituent has a long history. A  success of the 
Wounded Nucleon Model in the description of general characteristics of 
hadron - nucleus collisions in the wide energy range is remarkable. A more 
detailed study, \eg strange particle production, does not follow the 
strict WNM predictions.

 For heavier colliding objects, the Model fails. 
Attemps at a 
parametrization in terms of wounded quark participants seems to simplify 
the picture (but not to explain it!).

 Obviously, 
the central heavy ion collisions are not a simple superposition of 
nucleon - nucleon collisions. At the same time, the observed dominance of
`constituent scaling' in the production of particles from very different  
colliding objects, at very different energies, remains a puzzle.

\section {Acknowledgements}
I would like to express my cordial thanks to Barbara Wosiek, for numerous 
enlightening discussions, in particular on the PHOBOS data, and helping me 
in securing many experimental plots.

This work was partially supported by the Polish State Committee for 
Scientific Research grant 1 P03B 006 30.

\begin{figure}[p]
  \centering
  \includegraphics[width=0.95\textwidth]{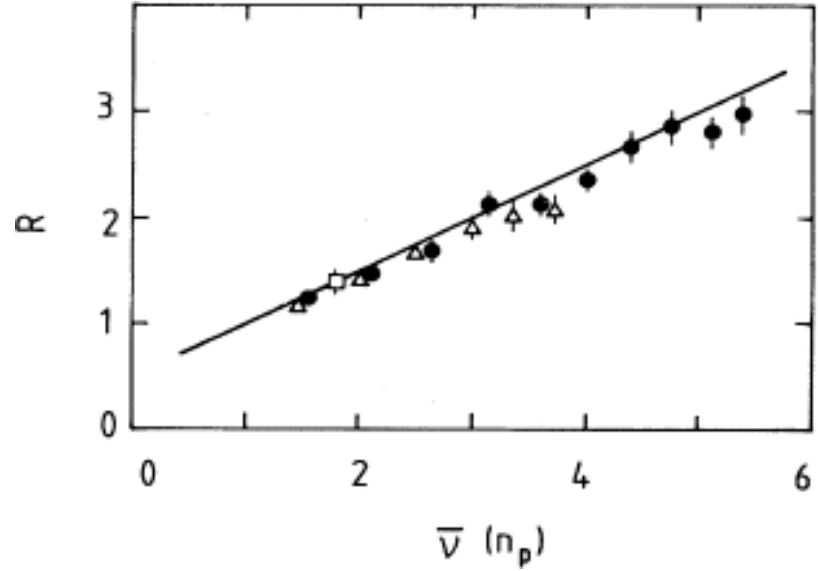}
  \caption{The ratio of charged particle multiplicity in p - A collisions to that in 
p - p collisions vs the average number $\nu$ of projectile collisions. A 
line shows  eq.~(1). (Ref.~3)}
  \label{fig:1}
\end{figure}

\begin{figure}[p]
  \centering
  \includegraphics[width=0.95\textwidth]{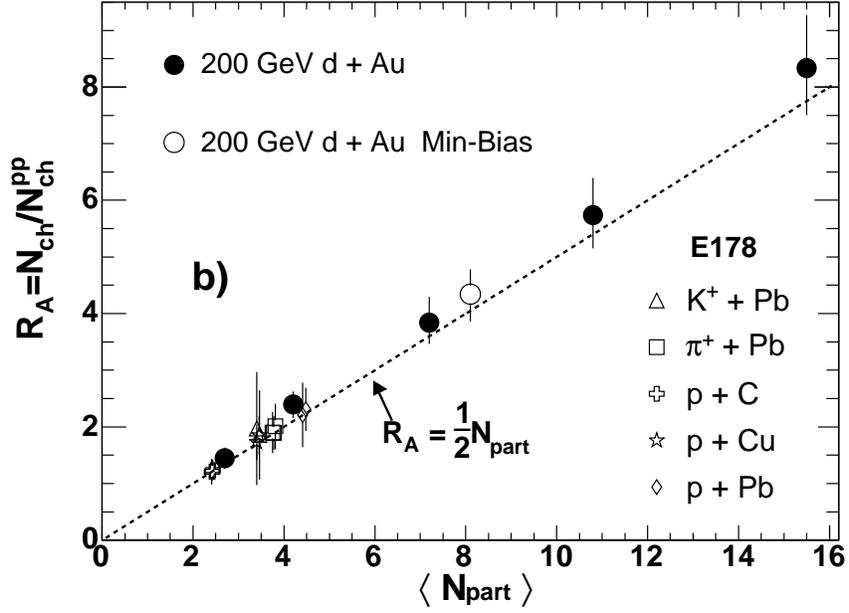}
  \caption{The ratio of the total charged particle multiplicity in hadron - nucleus 
and deuteron - nucleus collisions to that in hadron - proton collisions at 
the same energy, as a function of the number of participants. (Ref.~2)}
\end{figure}

\begin{figure}[p]
  \centering
  \includegraphics[width=0.95\textwidth]{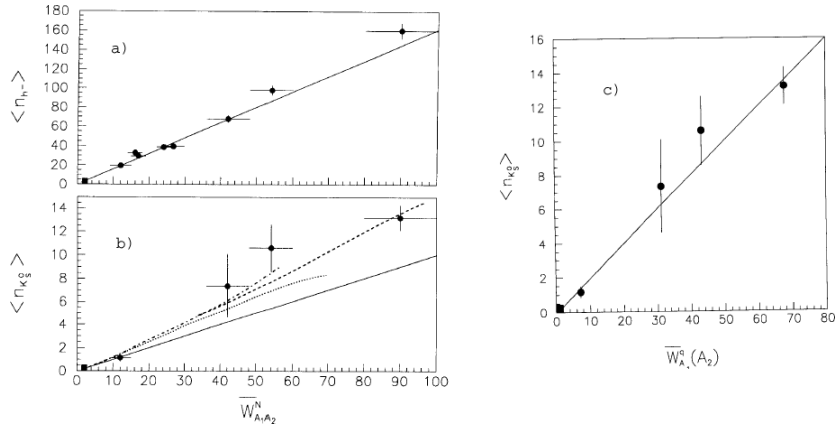}
  \caption{The average negative hadron multiplicity (a) and neutral kaon multiplicity 
(b) from A1 - A2 collisions at 200 GeV/N vs the number of nucleon 
participants. Part (c) shows the kaon multiplicity vs the number of 
wounded quarks. (Ref.~9)}
\end{figure}

\begin{figure}[p]
  \centering
  \includegraphics[width=0.95\textwidth]{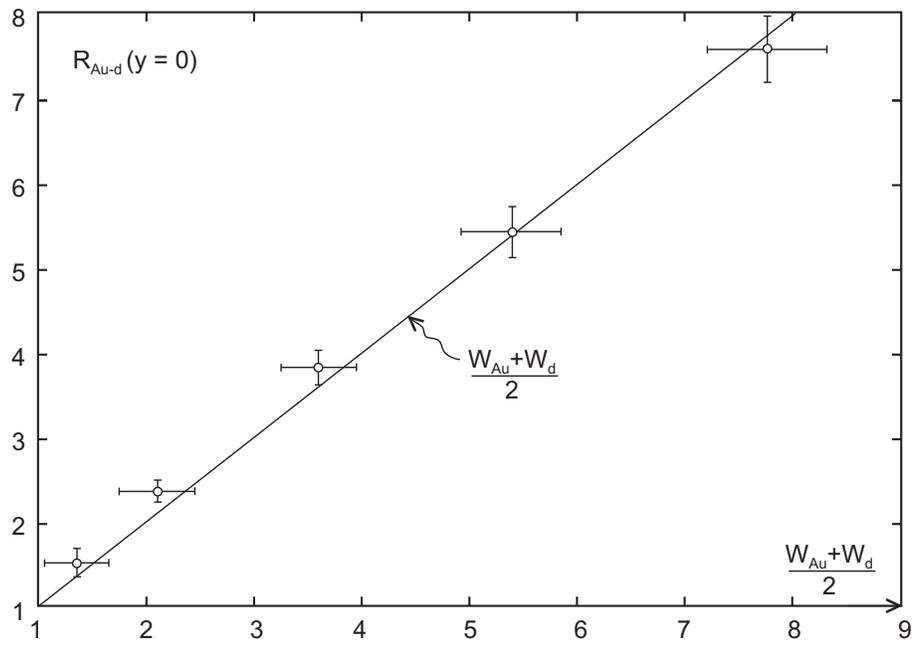}
  \caption{Charged particle multiplicity at midrapidity for deuteron - gold 
collisions at 200GeV/N compared with the predictions of the Wounded 
Nucleon Model. (Ref.~10)}
\end{figure}

\begin{figure}[p]
  \centering
  \includegraphics[width=0.95\textwidth]{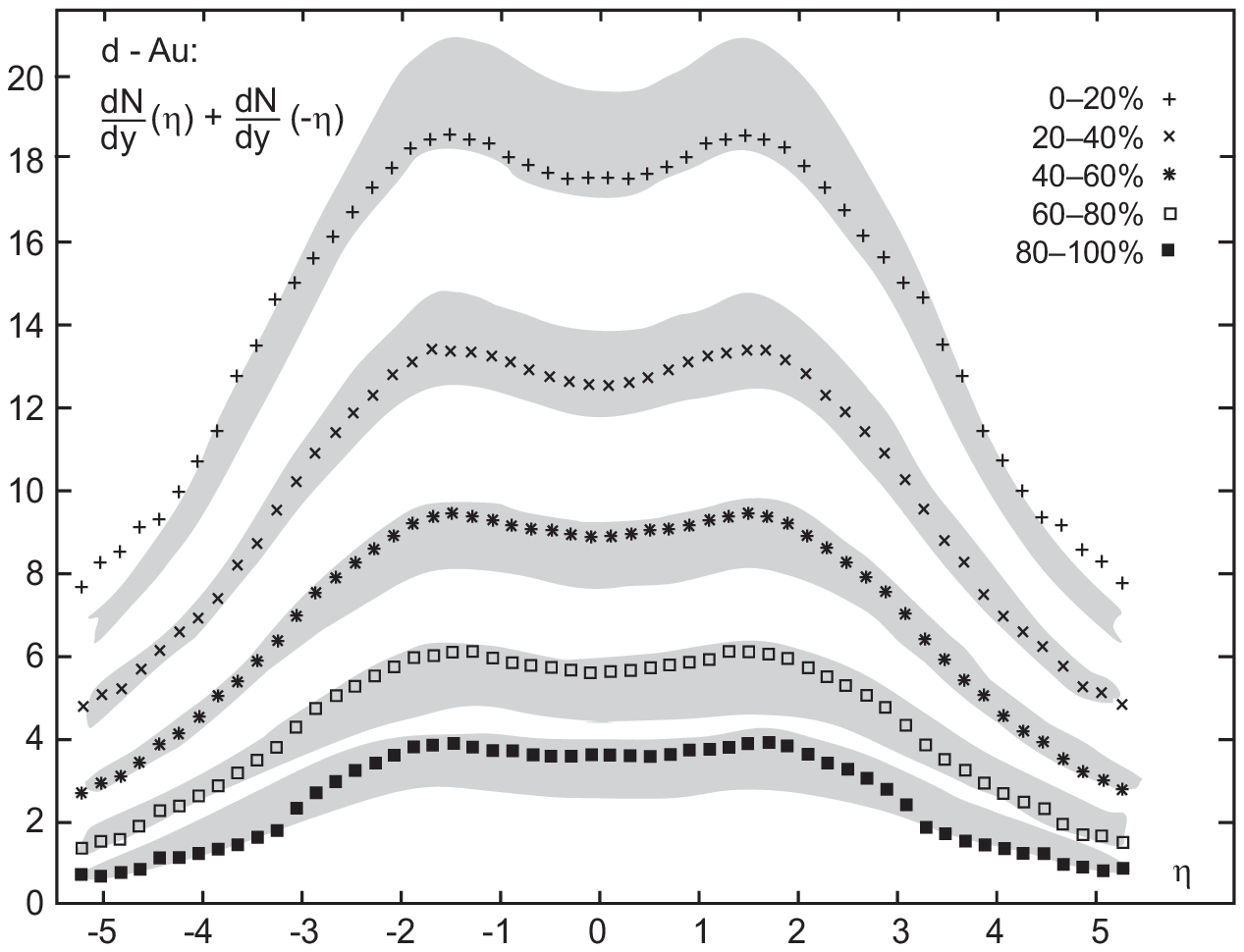}
  \includegraphics[width=0.95\textwidth]{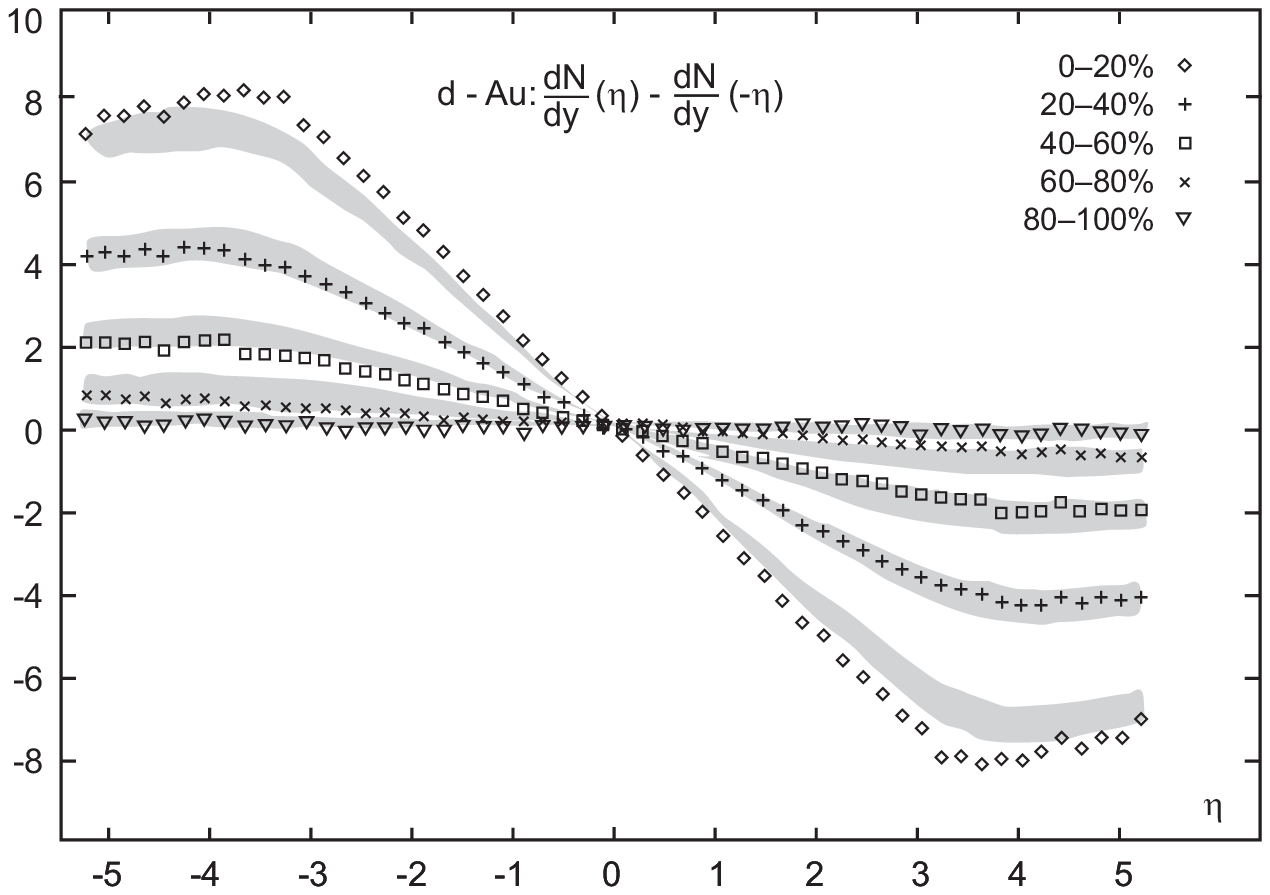}
  \caption{Symmetric and antisymmetyric part of the d - Au inclusive cross section 
compared with the predictions of the Wounded Nucleon Model. (Ref.~10)}
\end{figure}

\begin{figure}[p]
  \centering
  \includegraphics[width=0.95\textwidth]{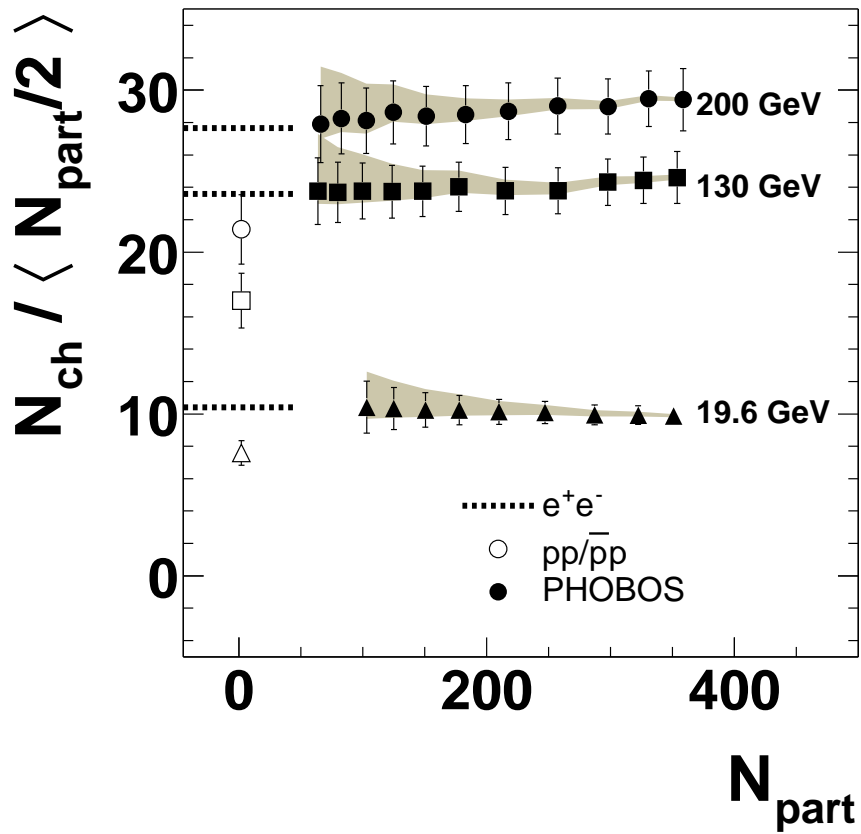}
  \caption{Total charged particle multiplicity per participant pair for Au - Au 
collisions at RHIC energies, vs the number of nucleon participants. 
Notice the p - p points lying systematically lower. (Ref.~13)}
\end{figure}

\begin{figure}[p]
  \centering
  \includegraphics[width=0.95\textwidth]{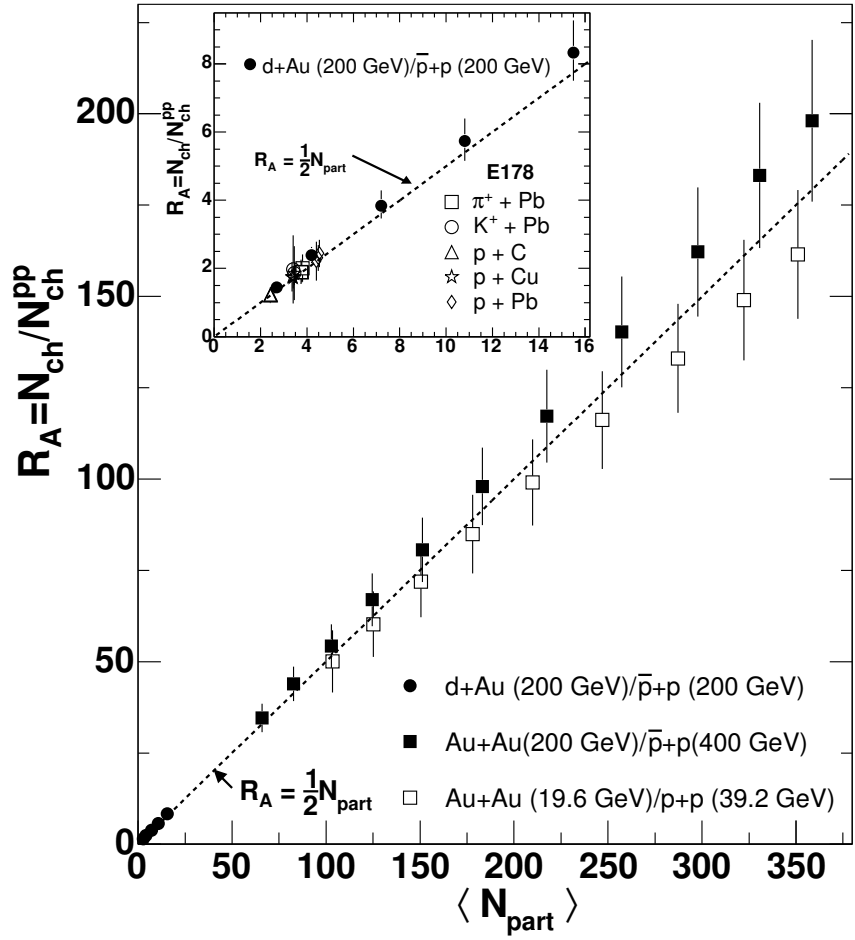}
  \caption{Ratios of the total particle multiplicity for nucleus - nucleus collisions 
for several energies, over the multiplicity in p - p collisions, vs the 
number of participating nucleons. For Au - Au interactions the p - p data 
is taken at twice the cms energy. (Ref.~13)}
\end{figure}

\begin{figure}[p]
  \centering
  \includegraphics[width=0.95\textwidth]{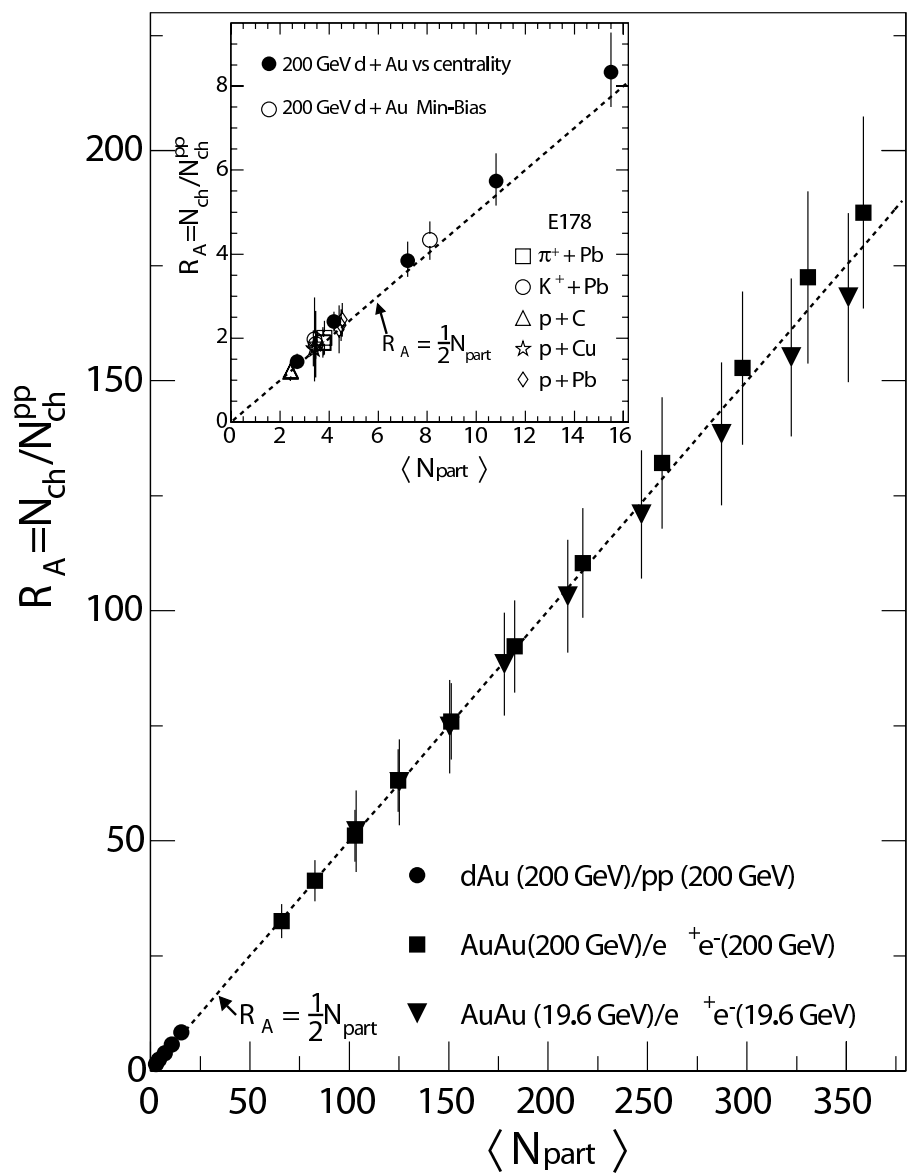}
  \caption{Same as Fig.~7, but the Au - Au data points are normalized to the 
multiplicity from e+e- collisions at the same energy. (Ref.~14)}
\end{figure}

\begin{figure}[p]
  \centering
  \includegraphics[width=0.85\textwidth]{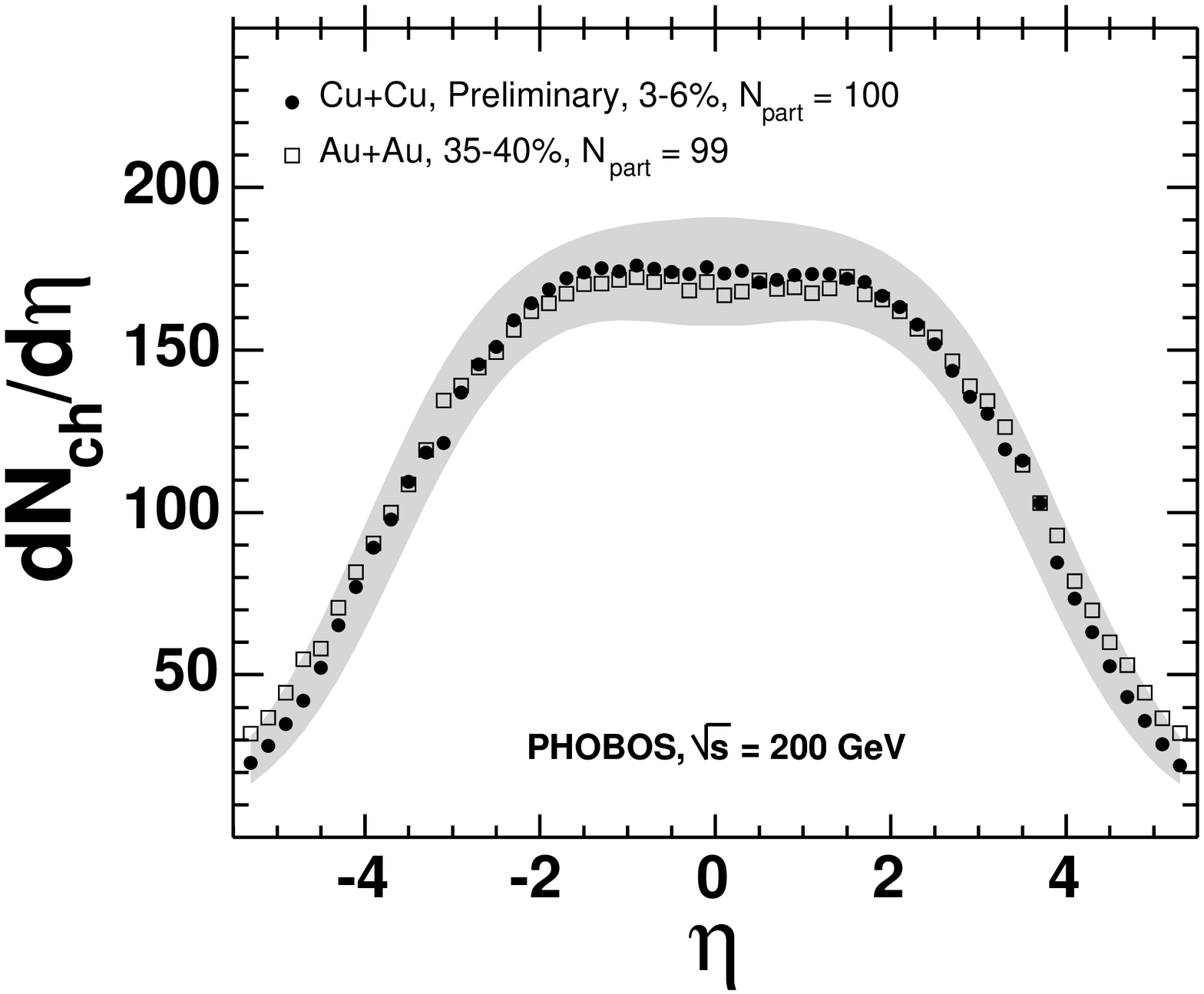}
  \includegraphics[width=0.85\textwidth]{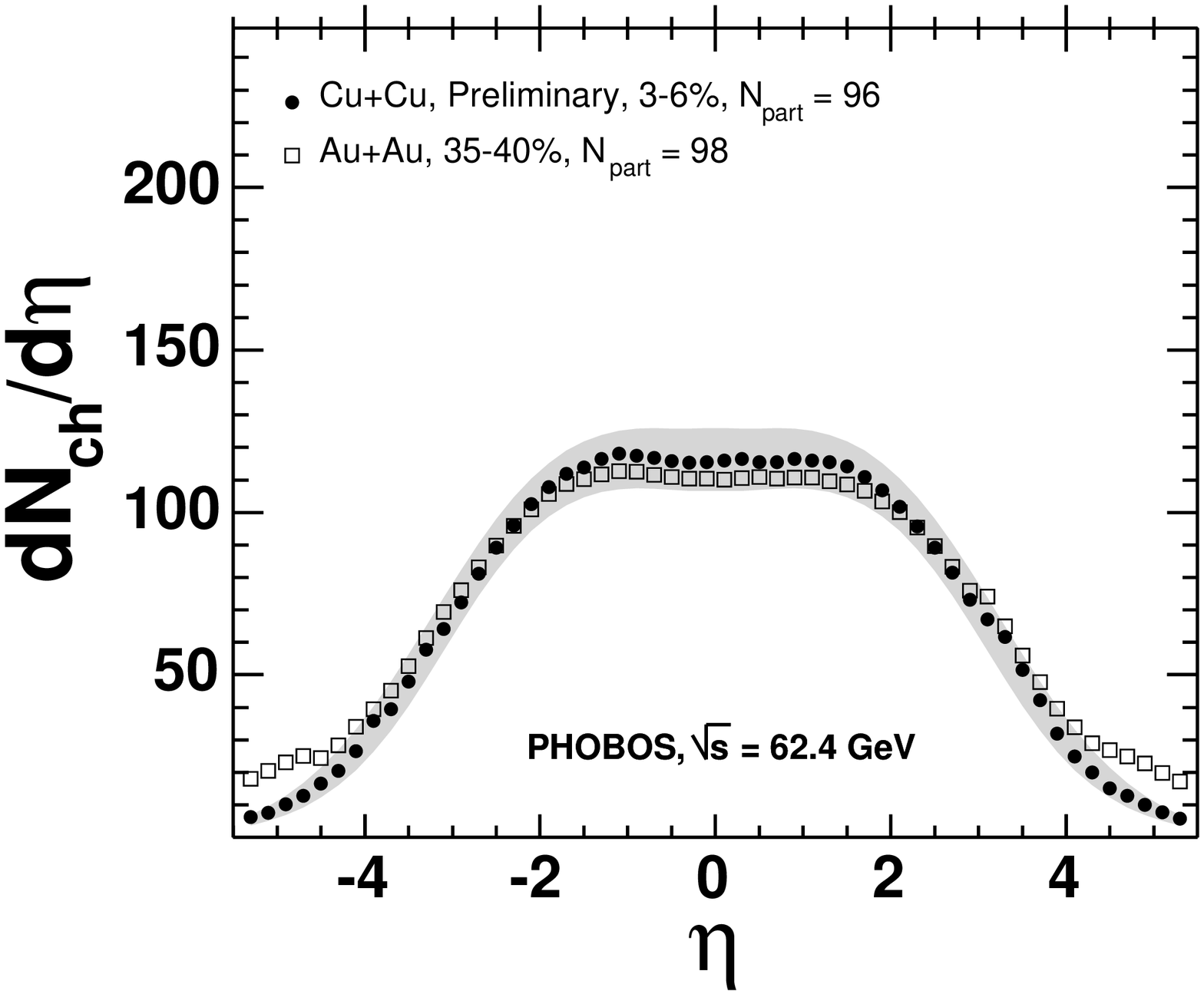}
  \caption{Charged particle pseudorapidity distributions for Cu - Cu and Au - Au 
collisions, measured at the same centrality (number of nucleon 
participants). (Ref.~15)}
\end{figure}

\begin{figure}[p]
  \centering
  \includegraphics[width=0.95\textwidth]{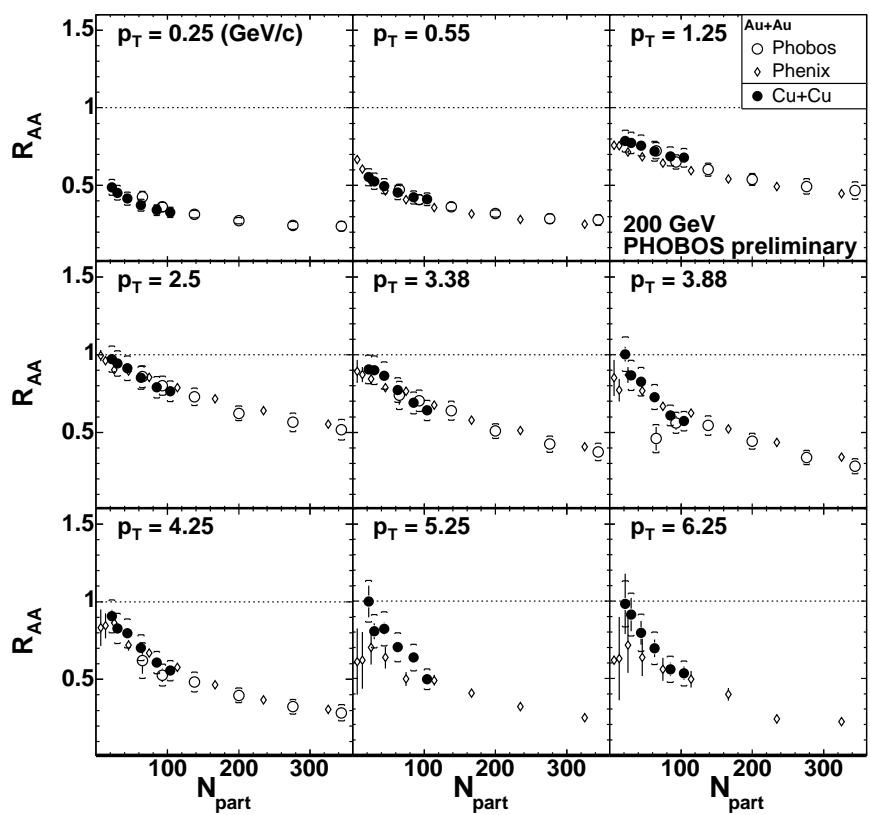}
  \caption{Nuclear modification factor in bins of pt, vs Npart for Cu - Cu and Au - 
Au at $\sqrt{s}$ $200 \,\mathrm{GeV}/N$. (Ref.~16)}
\end{figure}

\begin{figure}[p]
  \centering
  \includegraphics[width=0.95\textwidth]{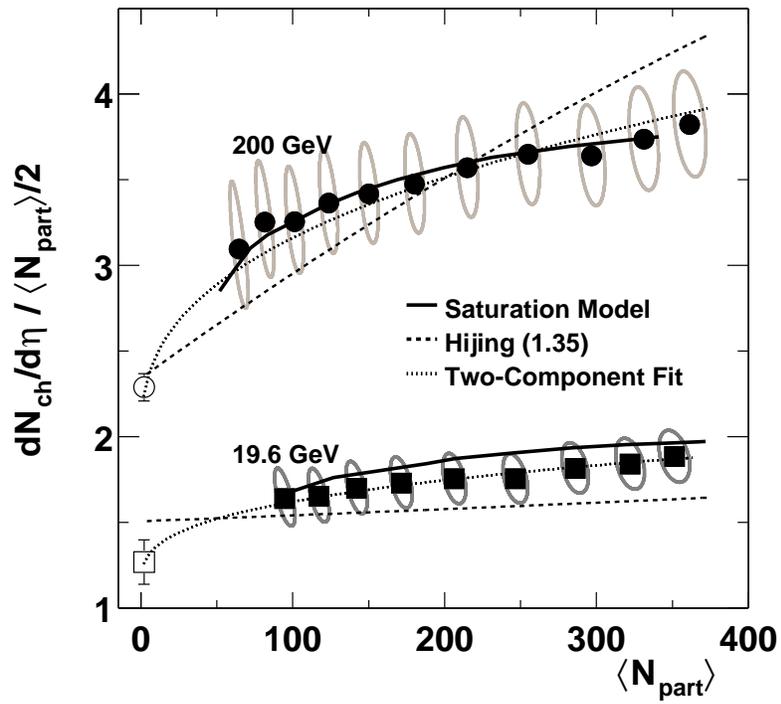}
  \caption{Midrapidity charged particle multiplicity per participant pair for Au - Au 
collisions vs the number of participants. (Ref.~13)}
\end{figure}

\begin{figure}[p]
  \centering
  \includegraphics[width=0.95\textwidth]{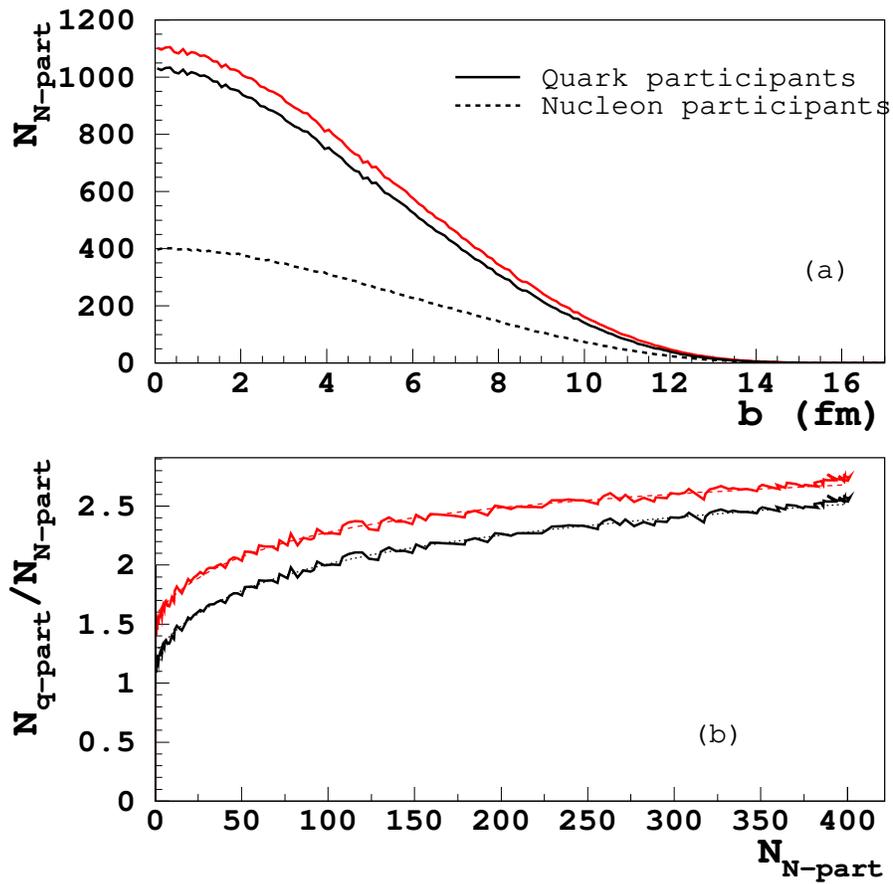}
  \caption{Calculated number of quark (solid line) and nucleon (dashed) participants 
vs the number of nucleon participants. Two sets of lines correspond to two 
values of quark-quark cross section. (Ref.~17)}
\end{figure}

\begin{figure}[p]
  \centering
  \includegraphics[width=0.95\textwidth]{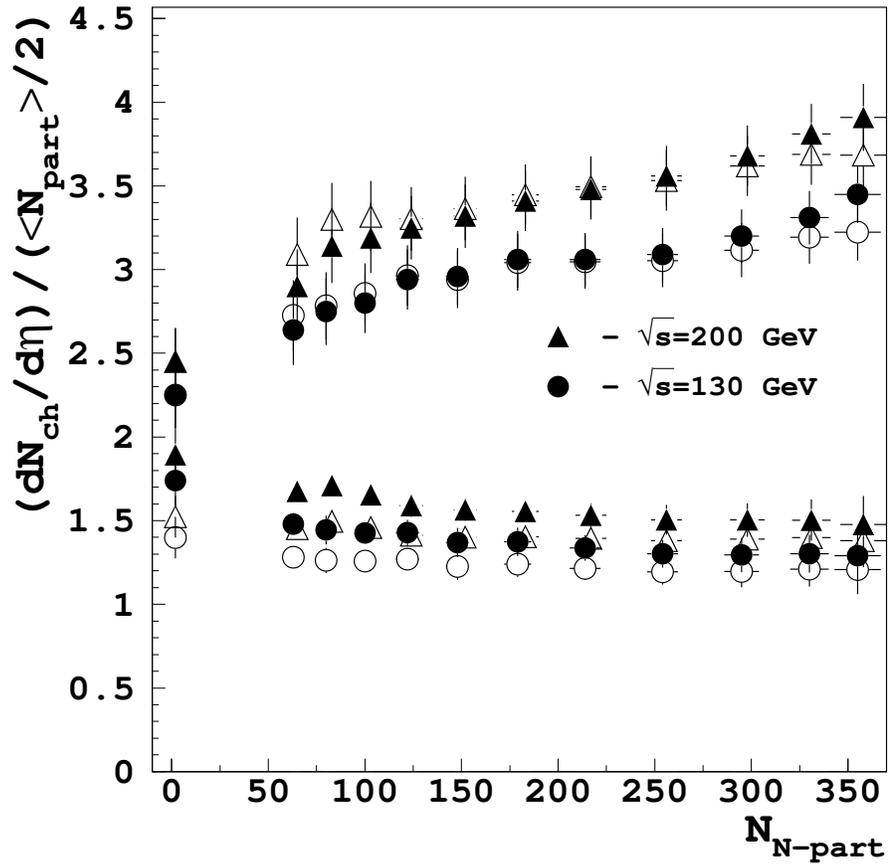}
  \caption{Midrapidity charged particle multiplicity per nucleon (upper) and quark 
(lower) participant pair, vs.centrality. Results for quark participants 
are shown for $\sigma_{qq}$ 4.5~mb (solid) and 6~mb (open symbols). (Ref.~17)}
\end{figure}

\begin{figure}[p]
  \centering
  \includegraphics[width=0.95\textwidth]{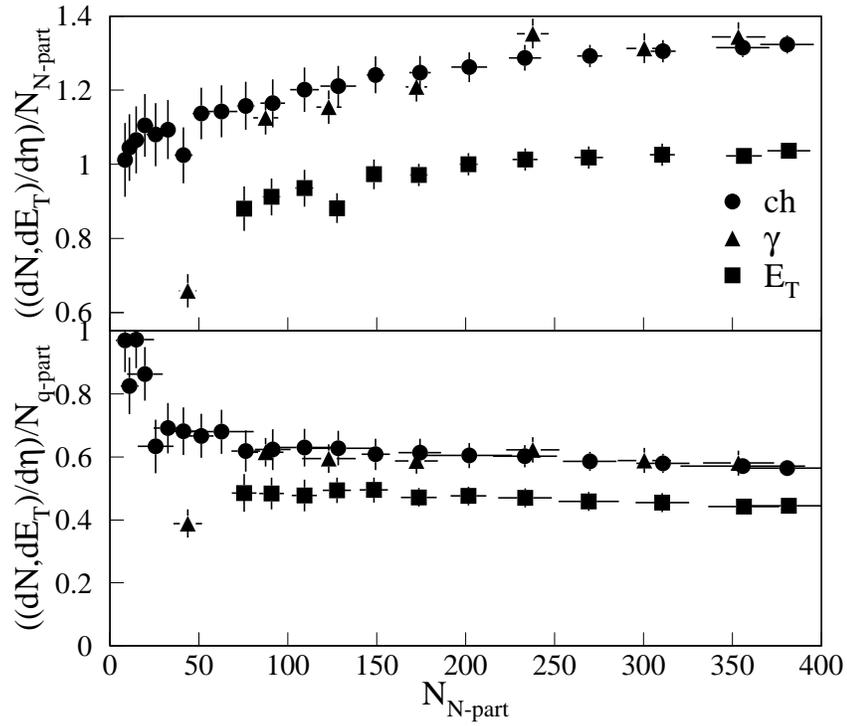}
  \caption{Midrapidity charged particle, gamma quanta and transverse energy density 
from Pb - Pb collisions at SPS, per nucleon (upper) and quark (lower) 
participant vs the number of nucleon participants. (Ref.~19)}
\end{figure}

\begin{figure}[p]
  \centering
  \includegraphics[width=0.95\textwidth]{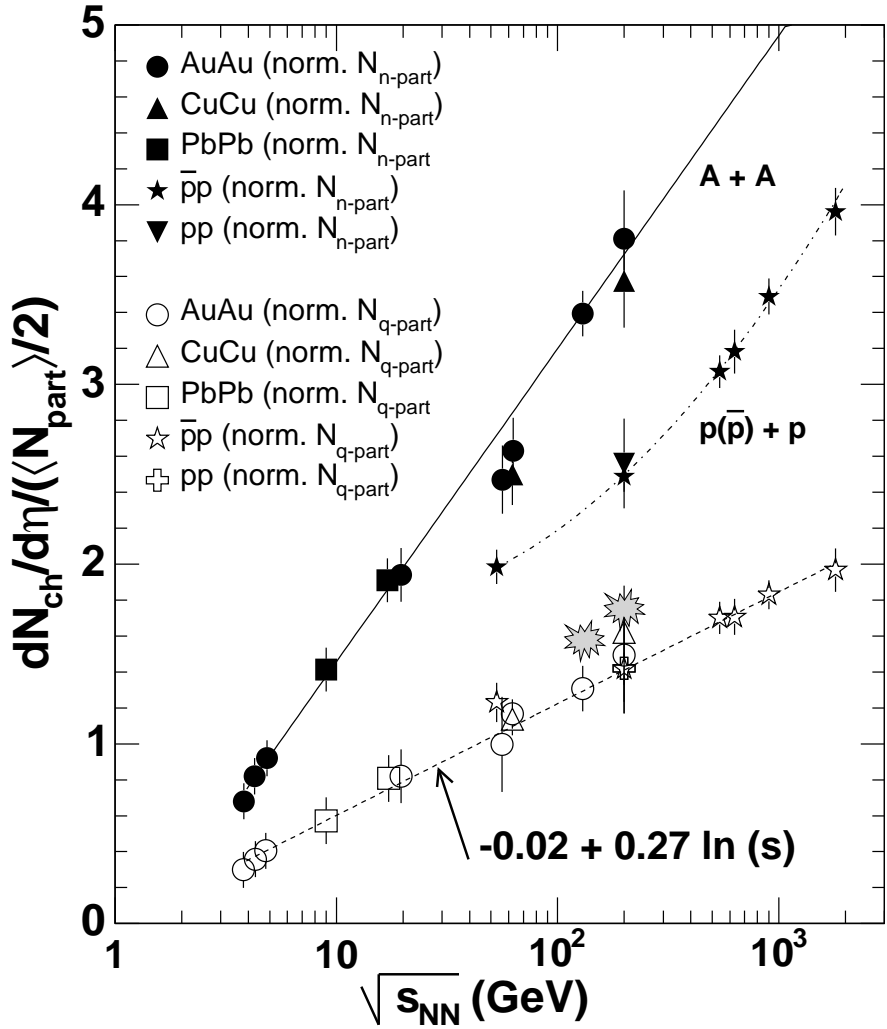}
  \caption{Particle density per quark (open symbols) an nucleon (solid points) 
participant pair for central nucleus - nucleus coillisions as a function 
of energy, and for proton - proton collisions. Two large points are 
calculated with number of quark participants from minimum bias pp 
collisions. (Ref.~23)}
\end{figure}

\end{document}